\documentclass[a4paper,fleqn,review]{cas-dc}
\usepackage[numbers]{natbib}
\usepackage{threeparttablex, booktabs, url, amsmath, xr, nature_abbr, flushend} %
\usepackage{lineno}
\usepackage{etoolbox}

\def\tsc#1{\csdef{#1}{\textsc{\lowercase{#1}}\xspace}}
\tsc{WGM}
\tsc{QE}

\usepackage{ulem}
\begin{document}
	\shorttitle{Zhu \& Zheng}
	\title [mode = title]{Superfluid Angular Momentum Reservoir Effect in Pulsar Glitches and Crab Pulsar Glitch Time Prediction} 
	\author[1]{Pei-Xin Zhu} %
	\author[2,3]{Xiao-Ping Zheng} %
	\cormark[1]
	\ead{zhxp@ccnu.edu.cn}
	\author[2]{Quan Cheng}
	\author[2]{Chenghui Niu}
	\author[4]{Erbil G\"{u}gercino\u{g}lu}
	
	\affiliation[1]{organization={School of Physics},addressline={Huazhong University of Science and Technology}, 
		city={Wuhan},
		citysep={}, %
		postcode={430074}, 
		country={China}}
	
	\affiliation[2]{organization={Institute of Astrophysics},
		addressline={Central China Normal University}, 
		city={Wuhan},
		citysep={}, %
		postcode={430079}, 
		country={China}}
	
	\affiliation[3]{organization={Department of Astronomy},
		addressline={Huazhong University of Science and Technology}, 
		city={Wuhan},
		citysep={}, 
		postcode={430079}, 
		country={China}}
	
	\affiliation[4]{organization={School of Arts and Science},
		addressline={Qingdao Binhai University}, 
		city={Qingdao},
		citysep={}, 
		postcode={266525}, 
		country={China}}
	
\cortext[1]{Corresponding author}

\maketitle

\begin{abstract}
Pulsar glitches are usually regarded as stochastic, independent events triggered by sudden angular momentum transfer from the neutron star’s superfluid interior to its crust. However, dense glitching episodes in the Crab pulsar suggest that some temporally proximate small glitches may instead form parts of broader dynamical episodes. Here we reanalyse more than five decades of Crab timing data by grouping nearby glitches into glitch clusters. In this clustered sequence, adjacent waiting times are consistent with preferred temporal organization around $\sim 3.5$ yr, and every-other cluster intervals indicate a longer-timescale component near $\sim 7$ yr. Cluster size correlates more strongly with preceding than with subsequent waiting times, with the clearest signal arising from the longer pre-history of the system. These results suggest that clustering primarily regularizes the temporal structure of the Crab glitch record and support a picture in which Crab glitches are better interpreted as temporally coupled, history-dependent collective events rather than as fully independent stochastic occurrences.
\end{abstract}

\section{Introduction}
Pulsar glitches provide one of the most direct observational probes of neutron-star interior dynamics\cite{link1999pulsar,haskell2015models,antonopoulou2022pulsar,antonelli2023insights}. In many sources, glitch occurrence appears irregular, and their waiting-time statistics can often be described, at least phenomenologically, by stochastic or critical-state models\cite{bak1987self,melatos2008avalanche,fulgenzi2017radio}. Yet this is not the whole picture. A small but important subset of pulsars exhibits preferred waiting times or clear size--waiting-time relations, indicating that longer-timescale temporal organization can coexist with apparently erratic event sequences\cite{eya2017angular}. The key question, therefore, is not simply whether glitches are random or periodic, but whether the observed sequence encodes a deeper dynamical structure that is obscured when events are considered one by one\cite{zhu2024glitches}. 

The Crab pulsar is a particularly instructive case\cite{basu2022jodrell}. It is one of the youngest and best-monitored glitching pulsars, and its activity is dominated by numerous small glitches superposed on a much smaller number of large events\cite{lyne201545}. Globally, the Crab waiting-time distribution has often been treated as approximately exponential, suggesting no obvious preferred timescale at the level of the full sequence\cite{howitt2018nonparametric}. Locally, however, the activity is clearly non-uniform, with the dense glitching episode showing time-dependent clustering and the largest glitches separated by intervals of roughly seven years\cite{carlin2019temporal,vivekanand2017non}. This combination of apparent short-timescale irregularity and possible long-timescale organization makes the Crab an important test case for whether glitch activity should be interpreted solely as a sequence of independent events.

A key difficulty is that, in the Crab, temporally proximate small glitches may not represent fully independent physical episodes. Treating every glitch as an isolated event may therefore obscure longer-timescale patterns that are distributed across groups of nearby events. This motivates a cluster-based description, in which closely spaced small glitches are combined into glitch clusters and analysed as components of a collective episode. From this perspective, the relevant statistical unit is not necessarily the individual glitch, but the clustered release pattern that emerges when short-separation events are considered together. 

Such a viewpoint naturally raises the possibility that Crab glitch activity is history-dependent. In this picture, successive observable glitches need not be statistically independent manifestations of a scale-free triggering process, but may instead form a temporally coupled sequence whose properties depend on the prior evolution of the system. We emphasize that this is an interpretive framework rather than a direct mechanistic proof. Its value lies in providing a way to test whether the seemingly irregular Crab sequence contains signatures of collective behaviour that become visible only when clustered events are treated as dynamically related.

In June 2025, before the subsequent Crab events, we placed a preliminary cluster-based forecast on the public record in an arXiv preprint, primarily to establish a time-stamped statement that could later be confronted with new observations\footnote{Zhu P X,Zheng X P. Normal distribution of crab pulsar glitch activity from a glitch cluster perspective, 2025, arXiv:2506.02925}. The two small glitches reported in July and August 2025 then provided the first external observational test of that preliminary framework and motivated the present reanalysis. Here we reanalyse the Crab glitch record from this cluster perspective using the updated sequence and a broader statistical assessment. We show that the waiting times between adjacent glitch clusters are consistent with organization around a characteristic timescale of approximately 3.5 years, and that the corresponding every-other cluster intervals indicate a longer-timescale component near 7 years. We further show that cluster size correlates more strongly with preceding than with subsequent waiting times, with the most robust signal arising from the longer pre-history of the system. Taken together, these analyses extend the earlier preliminary forecast into a broader statistical evaluation of the cluster framework, and support the view that Crab glitches are more naturally described as temporally coupled, history-dependent collective events than as a sequence of fully independent stochastic occurrences.

\begin{figure*}
	\centering
	\begin{minipage}{0.48\linewidth}
		\centering
		\includegraphics[width=\linewidth]{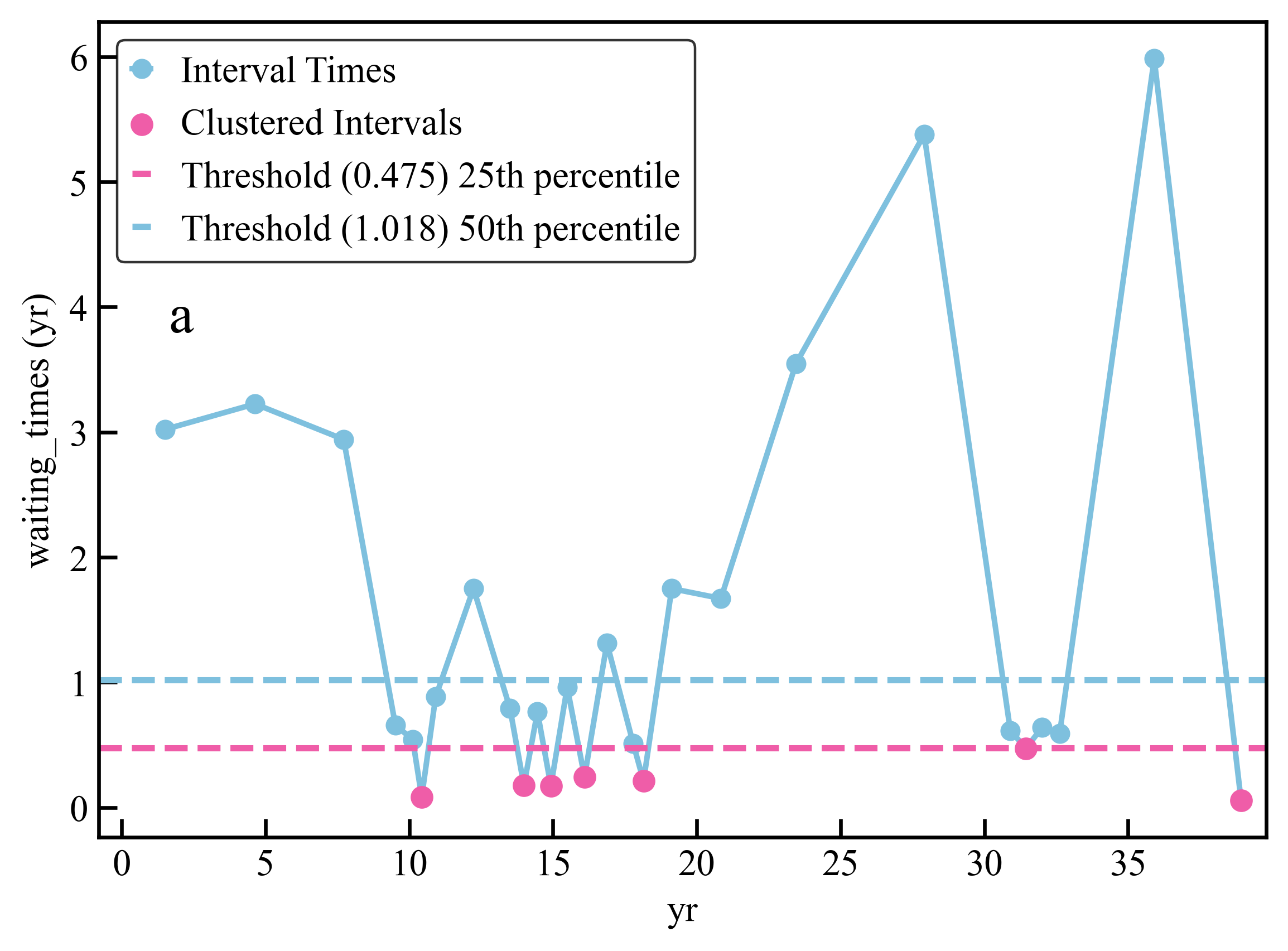}
	\end{minipage}
	\begin{minipage}{0.48\linewidth}
		\centering
		\includegraphics[width=\linewidth]{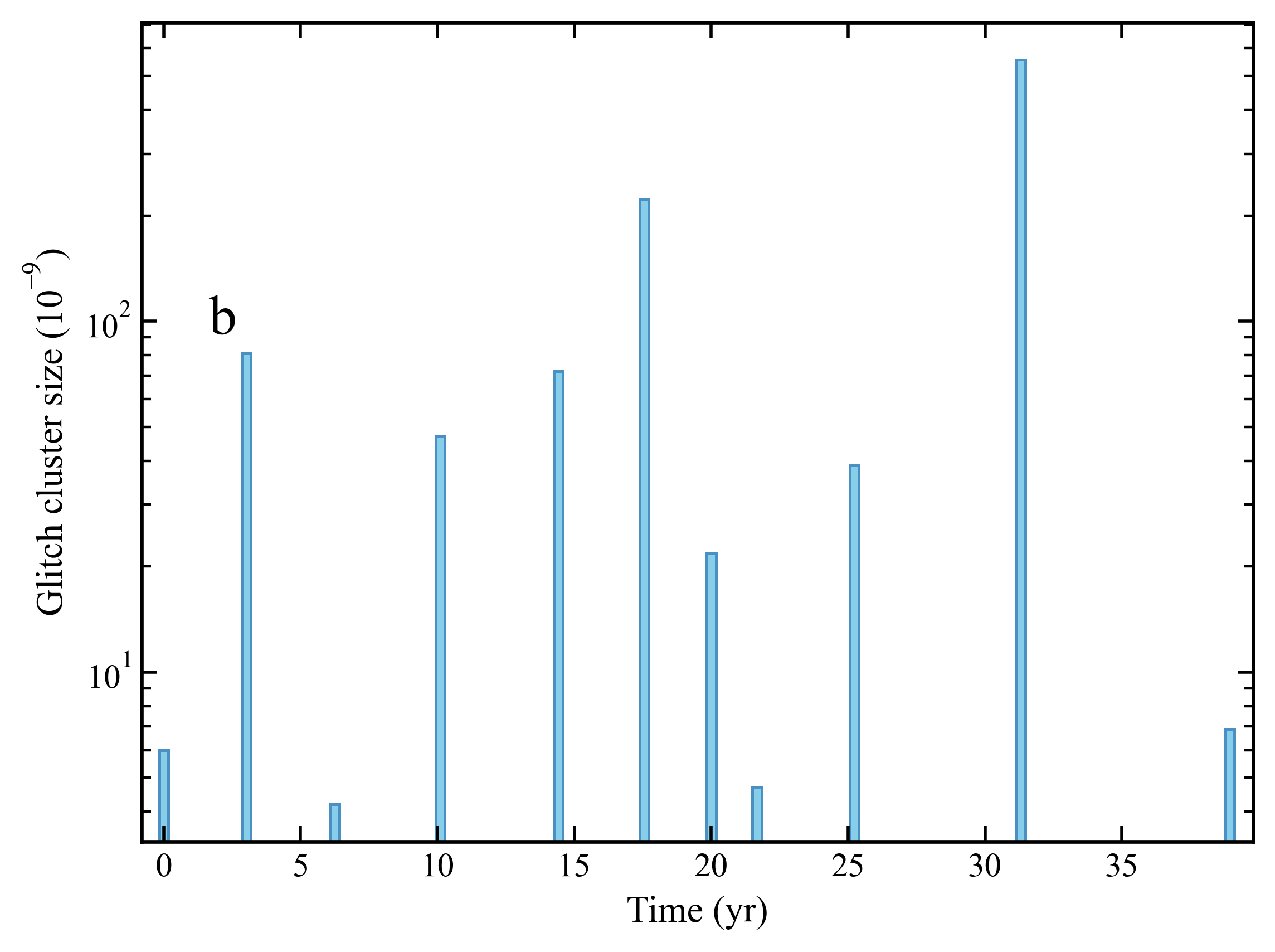}
	\end{minipage}	
	\caption{\textbf{| Threshold-based grouping of Crab glitches and the resulting glitch-cluster sequence.}
		\textbf{a}, Adjacent waiting times in the original 28-glitch sequence. The magenta points mark intervals identified as clustered intervals, namely those shorter than the 25th percentile of the fitted shifted-exponential waiting-time distribution. The dashed magenta and blue horizontal lines indicate the fitted 25th- and 50th-percentile thresholds, 0.475\,yr and 1.018\,yr, respectively. Intervals below the 25th percentile define dense regions, while the 50th-percentile threshold sets the temporal window used to merge temporally proximate glitches into clusters.
		\textbf{b}, Glitch-cluster sizes as a function of time after applying the threshold-based grouping procedure to the original Crab glitch sequence. Each bar represents one cluster, with the cluster size defined as the sum of the amplitudes of all glitches assigned to that group. The ordinate is shown on a logarithmic scale.
	\label{f1}}
\end{figure*}

\begin{figure*}
	\centering	
	\includegraphics[width=0.95\linewidth]{z3}
	\caption{\textbf{| Cumulative distribution functions of the glitch-cluster waiting times on the \(\sim 3.5\)-yr and \(\sim 7\)-yr timescales.}
		\textbf{a}, Empirical cumulative distribution function (ECDF) of the consecutive glitch-cluster waiting times, \(\Delta t^{(1)}\), corresponding to the \(\sim 3.5\)-yr timescale. The solid curve shows the best-fitting normal cumulative distribution function with \(\mu = 3.513\) yr and \(\sigma = 1.119\) yr.
		\textbf{b}, Same as in \textbf{a}, but for the every-other glitch-cluster waiting times, \(\Delta t^{(2)}\), corresponding to the longer \(\sim 7\)-yr timescale. The solid curve shows the best-fitting normal cumulative distribution function with \(\mu = 7.105\) yr and \(\sigma = 1.770\) yr.
		Together, the two panels show that the glitch-cluster waiting times are organized around characteristic timescales of approximately 3.5 yr and 7 yr, consistent with a longer-timescale quasi-periodic modulation in the Crab pulsar glitch activity.}
	\label{f2}
\end{figure*}

\section{Results}
\subsection{Quasi-periodicity}
A concentration of small glitches between approximately the 10th and 18th years is evident from the glitch-size evolution and inter-glitch intervals shown in Fig.~\ref{f1}. Previous studies also noted that this part of the Crab sequence is more clustered than expected from a uniform random occurrence of glitches \cite{lyne201545,carlin2019temporal}. This motivates a cluster-based description, in which temporally proximate small glitches are treated as components of a collective episode rather than as fully independent events.
	
Using the glitch clusters defined in Methods, we construct the primary waiting-time sequence, $\Delta t^{(1)}$, from the intervals between adjacent glitch clusters. When the sequence is viewed in this form, the cluster spacings appear substantially more organized than in the original event list, suggesting a preferred timescale on the order of a few years. To quantify this, the waiting-time distribution was assessed using the standard Anderson--Darling (AD) empirical-distribution-function statistic for normality, which assigns greater weight to tail deviations than the Cram\'er--von Mises statistic; the precise test definition and its Monte Carlo calibration are described in Methods. For the primary sequence, $\Delta t^{(1)}$, defined by intervals between adjacent glitch clusters, the observed AD statistic is 0.35, below the 5\% critical value of 0.68, indicating that normality cannot be rejected. A Monte Carlo analysis based on $10^4$ realizations of the fitted normal model gives a non-significant empirical $p$-value ($p \approx 0.42$), showing that the observed deviation is consistent with statistical fluctuations under the null hypothesis. 
	
This inference remains stable under a range of perturbations to the dataset. Excluding the maximum, the minimum, or both leaves the AD statistic below the corresponding critical threshold in all cases, with non-significant empirical p-values throughout. Likewise, a leave-one-out analysis gives AD statistics in the range 0.24-0.52, with none exceeding the corresponding critical value. Thus, despite the limited sample size, the clustered waiting-time sequence remains statistically consistent with a Gaussian description. Details of these robustness tests are provided in Supplementary Section~\ref{s2}.
	
The inclusion of the two most recent small glitches reported in ATel $\#$17298\footnote{\url{https://www.astronomerstelegram.org/?read=17298}} (2025 July 17) and ATel $\#$17331\footnote{\url{https://www.astronomerstelegram.org/?read=17331}} (2025 August 6) changes the fitted mean and dispersion only modestly relative to our preliminary June 2025 analysis, indicating that the inferred temporal organization is not driven by those new events alone. For the updated sample, the cumulative distribution function of $\Delta t^{(1)}$ is well described by a normal distribution with mean $\mu = 3.513$ yr and standard deviation $\sigma = 1.119$ yr, as shown in Fig.~\ref{f2}a. Taken together, these results indicate that the clustered Crab sequence is consistent with preferred temporal organization around a characteristic timescale of $\sim 3.5$ yr. 

\subsection{Longer-timescale organization near 7 years}
The clustered Crab sequence also shows evidence for temporal organization on a longer timescale. When the smallest-scale structure is suppressed, the three largest glitches are separated by intervals of 14.51 and 13.68~yr, suggesting an approximately 14-yr spacing among the largest events. In addition, intermediate glitch activity occurs near the midpoints of these long separations, implying that the clustered sequence may contain a corresponding component on a timescale of order 7~yr (Fig.~\ref{f1}). By the time of the July--August 2025 activity, 7.75~yr had elapsed since the 2017 large glitch, indicating that the most recent activity also occurs on a comparable longer timescale.
	
Motivated by this pattern, we constructed an alternative waiting-time sequence, $\Delta t^{(2)}$, from every-other glitch-cluster separations (1st--3rd, 2nd--4th, and so on), in order to probe temporal structure beyond adjacent cluster intervals. For this sequence, the Anderson--Darling (AD) statistic is 0.16, well below the 5\% critical value of 0.68, and Monte Carlo calibration gives a high empirical $p$-value ($p \approx 0.95$), indicating that the observed deviations are entirely consistent with statistical fluctuations under the fitted Gaussian model. This result is stable under perturbations of the sample: excluding the maximum, the minimum, or both leaves all AD statistics small ($\lesssim 0.21$) and all empirical $p$-values high ($p \gtrsim 0.8$), while leave-one-out realizations give AD statistics in the range 0.14-0.21, with none approaching the corresponding critical threshold. Thus, the longer-timescale sequence is not driven by any single observation.
	
Before the most recent glitches, the every-other cluster waiting times were described by a normal distribution with mean $\mu = 6.677$~yr and standard deviation $\sigma = 1.551$~yr. After incorporating the July and August 2025 activity through the same working construction adopted for the updated clustered sequence, the corresponding distribution is fitted by a normal model with mean $\mu = 7.105$~yr and standard deviation $\sigma = 1.770$~yr, as shown in Fig.~\ref{f2}b. The fitted location of this longer-timescale component is therefore only modestly shifted by the inclusion of the new data. Taken together, the every-other waiting times indicate that the clustered Crab sequence contains a longer-timescale component $\sim7$~yr, complementing the $\sim3.5$~yr organization seen in adjacent cluster intervals. This longer-timescale structure is qualitatively consistent with the approximately 14-yr spacing of the largest glitches and with the appearance of intermediate activity near the half-period points.
	
To test whether this organization requires a preferred timescale, we compared Gaussian, bounded uniform, and exponential descriptions for both the adjacent and every-other waiting-time sequences. In both cases, the exponential model is strongly disfavoured by information-criterion comparisons and parametric-bootstrap goodness-of-fit tests, arguing against a simple memoryless description of the clustered Crab activity. By contrast, both Gaussian and bounded uniform models remain statistically admissible for the present small samples, showing that Gaussian consistency alone does not uniquely establish quasi-periodicity. For this reason, our interpretation does not rely on marginal distribution fitting alone. Instead, the evidence for preferred temporal organization in the Crab glitch sequence comes from the combined presence of clustered $\sim3.5$-yr intervals, a corresponding longer-timescale component near $\sim7$~yr, and the non-trivial midpoint concentration of smaller clusters between successive large glitches (Supplementary Fig.~\ref{f7}, \ref{f9}, Supplementary Tables~\ref{t2}, and Supplementary Note~\ref{s3}, \ref{s6}).

\begin{figure*}
	\centering	
	\includegraphics[width=0.95\linewidth]{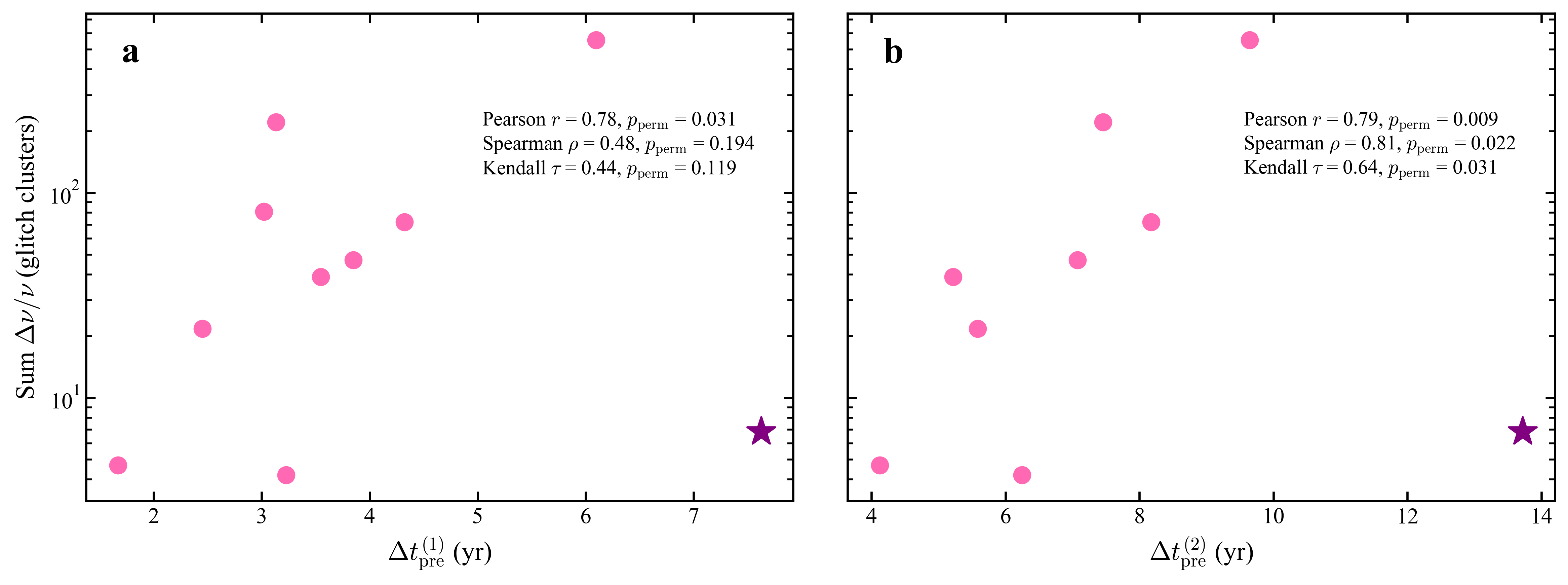}
	\caption{\textbf{| Cluster size versus preceding waiting time on adjacent and two-step timescales.}
		\textbf{a}, Cluster size as a function of the pre-cluster waiting time, $\Delta t^{(1)}_{\mathrm{pre}}$, defined from adjacent glitch clusters and associated with the $\sim 3.5$-yr timescale. A positive trend is present, indicating that larger clusters tend to occur after longer immediately preceding intervals, although the rank-based support is more modest than the Pearson trend alone would suggest.
		\textbf{b}, Cluster size as a function of the pre-two-step waiting time, $\Delta t^{(2)}_{\mathrm{pre}}$, defined by spanning the two preceding clusters and associated with the longer $\sim 7$-yr timescale. This relation provides the most internally consistent correlation among the five tested waiting-time constructions, indicating that the strongest size--waiting-time signal is linked to the longer pre-history of the system. In both panels, the most leveraged point is indicated separately to show that the stronger pre-two-step relation is not a generic consequence of a single outlier.}
	\label{f3}
\end{figure*}
\subsection{Cluster size and waiting times}
Beyond the waiting-time distributions themselves, the cluster framework allows a further test of history dependence: whether the size of a glitch cluster is linked more strongly to the system's prior waiting-time history than to its subsequent evolution. To examine this, we consider five waiting-time constructions. For adjacent clusters, we define the pre- and post-cluster intervals, $\Delta t^{(1)}_{\mathrm{pre}} = t_i - t_{i-1}$ and $\Delta t^{(1)}_{\mathrm{post}} = t_{i+1} - t_i$. To probe longer-baseline correlations, we also define three two-step intervals spanning one intervening cluster: $\Delta t^{(2)}_{\mathrm{pre}} = t_i - t_{i-2}$, $\Delta t^{(2)}_{\mathrm{mid}} = t_{i+1} - t_{i-1}$, and $\Delta t^{(2)}_{\mathrm{post}} = t_{i+2} - t_i$.

These five constructions are designed to test different forms of temporal coupling. The adjacent pre- and post-cluster intervals ask whether cluster size is linked more strongly to the immediately preceding buildup or to the subsequent waiting time after the release. The two-step quantities extend this idea to a longer baseline. In particular, $\Delta t^{(2)}_{\mathrm{pre}}$ tests whether the observable cluster size depends on a more extended pre-history than is captured by the nearest-neighbour interval alone, whereas $\Delta t^{(2)}_{\mathrm{post}}$ probes whether any comparable influence extends into the longer subsequent evolution. By contrast, $\Delta t^{(2)}_{\mathrm{mid}} = t_{i+1}-t_{i-1}$ measures the total two-interval window centred on the $i$th cluster, and therefore tests whether the relevant timescale is distributed more symmetrically around the event rather than being preferentially associated with the pre-history.

\begin{figure}
	\centering
	\includegraphics[width=1.0\linewidth]{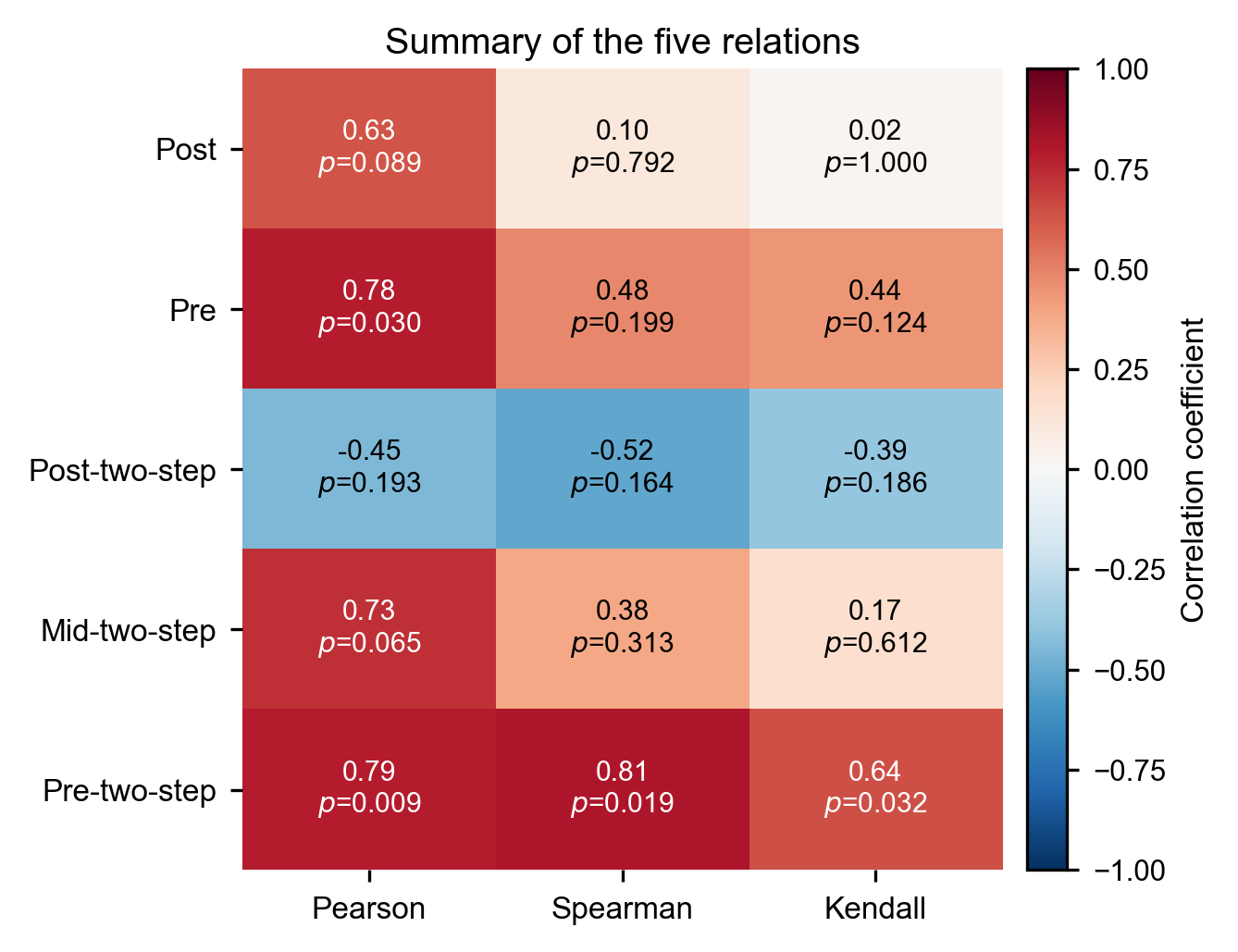}
	\caption{\textbf{| Summary of cluster-size correlations for five waiting-time constructions.}
		Heatmap of the Pearson, Spearman and Kendall correlation coefficients between cluster size and the five waiting-time definitions: post, pre, post-two-step, mid-two-step and pre-two-step. Numerical values and permutation-calibrated $p$-values are shown in each cell. Two features are apparent. First, correlations involving preceding waiting times are systematically stronger than those involving subsequent waiting times. Second, the pre-two-step relation is the only one that remains strong across all three correlation measures, indicating that the most robust signal is associated with the longer pre-history of the clustered sequence. By contrast, the mid-two-step and post-two-step constructions show weaker or less consistent support, implying that the correlation is not simply controlled by a symmetric multi-interval window around each event or by the subsequent evolution alone.}
	\label{f4}
\end{figure}

A clear directional pattern emerges. Across the five waiting-time definitions, correlations with preceding waiting times are systematically stronger than those with subsequent waiting times. On the adjacent-cluster timescale, cluster size increases with the pre-cluster waiting time (Fig.~\ref{f3}a), whereas the corresponding post-cluster relation is weak and unstable. On the longer two-step timescale, the strongest signal is obtained for the pre-two-step interval (Fig.~\ref{f3}b). In the summary heatmap (Fig.~\ref{f4}), the pre-two-step relation is the only one that remains strong across Pearson, Spearman and Kendall measures simultaneously, whereas the mid-two-step and post-two-step constructions do not show comparably robust support. The main statistical result of this section is therefore not a generic size-waiting-time correlation, but a directional asymmetry: cluster size is linked more strongly to pre-history than to post-history, and most strongly to the longer pre-history traced by $\Delta t^{(2)}_{\mathrm{pre}}$.

The adjacent pre-cluster relation indicates that larger clusters tend to occur after longer preceding intervals, although the rank-based support remains more modest than the Pearson trend alone would suggest. By contrast, the pre-two-step relation provides the most internally consistent result among the five constructions, indicating that the observable cluster size retains information from more than one preceding cluster interval. In this sense, the cluster sequence is not well described as a purely event-by-event process. Rather, the strongest correlation appears only when the waiting time is constructed to include a longer segment of the system's prior evolution. The weaker support for the mid- and post-two-step constructions further suggests that the correlation is not simply controlled by a symmetric multi-interval window around each event, nor by the subsequent evolution alone. Instead, the strongest signal is associated specifically with the longer pre-history of the system.

The treatment of the most recent cluster requires particular care. The two small glitches observed in July and August 2025 satisfy the temporal criterion for a cluster, but in the absence of any confirmed accompanying major glitch, the cumulative amplitude of that cluster remains uncertain. For this reason, data points involving the unresolved latest cluster are shown in the scatter plots for completeness but excluded from the corresponding correlation calculations. This precaution is especially important for the post-cluster relation, where an apparent positive trend is largely driven by leverage from the largest cluster and disappears once the unresolved endpoint is excluded. The pre-two-step relation remains the most robust even under this conservative treatment, whereas the mid-two-step construction should be interpreted more cautiously because it is more sensitive to the placement of the latest unresolved cluster.

The contrast with PSR J0537$-$6910 is also noteworthy. In J0537$-$6910, glitch magnitude is well known to correlate with the subsequent waiting time, a pattern commonly interpreted within a threshold-triggered framework \cite{middleditch2006predicting,antonopoulou2018pulsar}. In the Crab, by contrast, the clearer association is with preceding waiting times, particularly on the longer two-step baseline. We therefore interpret the Crab results more cautiously as evidence for directional, history-dependent temporal coupling in the clustered sequence.

\subsection{Prospective consistency of the June 2025 forecast}
In June 2025, before the July-August 2025 Crab events, we reported a preliminary cluster-based forecast in an earlier arXiv preprint, predicting that the next Crab glitches would occur before February 2026. The two small glitches subsequently reported on 17 July and 6 August 2025 are therefore consistent with that prospective forecast and provide an external observational test of the framework.

This prospective consistency motivated a broader reanalysis of the clustered sequence. In particular, the updated size--waiting-time correlations suggest that the recent minor activity is more naturally interpreted as part of an ongoing active episode than as the complete release of a larger cluster. Within this picture, the July-August 2025 glitches may represent precursor activity rather than the full expression of the longer-timescale clustered sequence. Quantitative extrapolations based on the adjacent and pre-two-step relations are given in Supplementary Section~\ref{s5}.

\subsection{CDF of the glitch cluster sizes}
To test whether the clustering procedure alters not only the waiting-time statistics but also the amplitude statistics, we inspected the cumulative distribution function (CDF) of the glitch-cluster sizes. In contrast to the clustered waiting times, which are statistically consistent with organization around characteristic long-timescale components, the cluster-size distribution does not show comparably strong evidence for a narrow characteristic scale. Instead, it remains broad and is qualitatively consistent with a heavy-tailed form. This indicates that the clustering procedure primarily reorganizes the temporal structure of the Crab glitch sequence, while leaving the amplitude statistics much less regularized. The emergence of a preferred waiting-time scale therefore does not imply a correspondingly sharp preferred scale in cluster size.

\section{Discussion}
Our results suggest that the Crab glitch record is more structured when temporally proximate events are treated as glitch clusters rather than as fully independent glitches. In this clustered sequence, the waiting times are statistically consistent with preferred temporal organization on both the adjacent-cluster scale ($\sim3.5$~yr) and the every-other-cluster scale ($\sim7$~yr), while the strongest size--waiting-time correlation is obtained not for subsequent intervals but for preceding ones, especially on the longer pre-two-step baseline. Taken together, these results favour a description in which the Crab sequence retains a measurable dependence on its prior evolution. This point is central to the cluster framework adopted here.
	
In the Crab pulsar, the abundance of small glitches and the presence of dense glitching episodes make it unlikely that every recorded event corresponds to an independent dynamical episode of equal statistical relevance. From this perspective, clustering is not merely a regrouping procedure, but a way of redefining the effective event unit so that temporally proximate small glitches are treated as components of a broader release pattern. Under this description, longer-timescale temporal organization becomes more apparent than in the raw event sequence, and the directional size--waiting-time correlations can be tested in a way that is not accessible at the level of individual glitches alone.
	
The directional asymmetry of the correlations is physically informative. In particular, cluster size is linked more strongly to preceding waiting times than to subsequent ones, and most strongly to the longer pre-two-step history. This indicates that the observable cluster size depends on more than the immediately preceding interval and is therefore more consistent with a history-dependent collective process than with a purely local event-by-event picture. The contrast with PSR J0537$-$6910 is also notable: in PSR J0537$-$6910, glitch magnitude is most clearly associated with the subsequent waiting time, whereas in the Crab the clearer association is with the preceding temporal history\cite{middleditch2006predicting,antonopoulou2018pulsar}. One possible interpretation is that these apparently different directional signatures do not necessarily imply entirely different underlying reservoir physics, but rather different observational expressions of it. Because the Crab pulsar is much younger than PSR J0537$-$6910, its observable glitch phenomenology may reflect a different stage of neutron-star evolution.
	
At the same time, the present results remain subject to important limitations. The clustered Crab sample is small, the waiting-time constructions are not strictly independent, and Gaussian consistency alone does not uniquely establish a strict periodic clock. Our midpoint Monte Carlo test argues against the simplest midpoint-coincidence explanation of the apparent $\sim3.5$~yr structure, although the result remains suggestive rather than definitive because the number of relevant intervals is still limited. Another notable feature is the asymmetry between temporal and amplitude statistics: clustering regularizes the waiting-time sequence much more strongly than the cluster-size distribution, which remains broad. This suggests that, in the Crab pulsar, the long-timescale reservoir effect may regulate when glitch activity occurs more strongly than how much observable spin-up is released in an individual clustered event. 
One possible explanation is that the observable glitch amplitude in the Crab does not directly trace the full amount of angular momentum or free energy accumulated prior to a clustered event. This possibility may be related to the young physical state of the Crab pulsar and to the persistent-shift phenomenon, in which the post-glitch braking torque does not fully return to its pre-glitch level \cite{lyne1992spin,ge2020discovery}. Internal dissipative channels, potentially including processes such as Cooper-pair breaking, may contribute to this mismatch, although the present work does not provide a quantitative demonstration of that mechanism\cite{page2011rapid,ho2012rotational,zheng2025persistent}. If so, a characteristic timescale may emerge in the waiting-time statistics even while the cluster-size distribution remains broad. Future monitoring of the current active phase, especially if the next major Crab glitch can be observed in full, will provide the clearest test of whether the present minor activity belongs to a broader clustered episode and will place tighter constraints on how the Crab pulsar stores and releases angular momentum over long timescales.

\section{Methods}
\subsection{Glitch dataset and time origin}
Because the Crab pulsar glitch record contains an observational gap between February 1979 and February 1982 \cite{lyne201545}, we adopted the first post-1982 glitch (MJD 46663.69) as the initial event of the analysed sequence and defined this epoch as the temporal origin. Under this convention, the present analysis includes 28 of the 32 recorded Crab glitches.

\subsection{Construction of glitch clusters}
To identify temporally proximate groups of glitches, we first computed the waiting times in the original 28-glitch sequence and fitted them with a shifted exponential model. The 25th and 50th percentiles of this fitted distribution were then used as operational thresholds for cluster construction. Intervals shorter than the 25th percentile were used to identify dense glitching regions, and the 50th percentile was used to define the temporal window within which neighbouring glitches were merged into a single cluster. Under this procedure, glitches 4--8, 9--15, 16--18, 22--26 and 27--28 were merged, while the remaining events retained their original temporal separation.

In fitting the original inter-glitch waiting times, we allowed the location parameter of the exponential model to vary rather than fixing it to zero. This choice reflects the fact that the observed glitch catalogue is unlikely to be complete at very short separations: strong post-glitch recovery can mask subsequent weak events, and very small glitches may remain unresolved under finite timing precision and cadence. We therefore interpret the fitted non-zero location parameter as an effective lower bound on resolvable separations in the observed catalogue, rather than as direct evidence for a strict intrinsic refractory timescale.

Although the 25th- and 50th-percentile thresholds are empirical, the resulting cluster assignments are stable because the Crab sequence contains both conspicuously dense small-glitch episodes and clearly extended intervals between the major events. To facilitate reproducibility, the original glitch list and all derived cluster quantities are provided in Supplementary Table~\ref{t1}.

\subsection{Cluster size and representative epoch}
For the \(i\)th glitch cluster, we define the cluster size as the sum of the amplitudes of its constituent glitches,
$$ \Delta \nu_i/\nu = \sum_{k \in i} \Delta \nu_k/\nu . $$
The representative epoch of the \(i\)th cluster is defined as the size-weighted mean epoch of its constituent glitches,
$$ t_i=\frac{\sum_{k \in i} (\Delta \nu_k/\nu)\, t_k}{\sum_{k \in i} (\Delta \nu_k/\nu)} . $$
This weighted-mean epoch was adopted so that the representative time reflects the effective timing of the total observable release within the cluster.

\subsection{Waiting-time constructions}
Using the representative cluster epochs \(t_i\), we constructed two primary waiting-time sequences for the timing analysis,
\[
\Delta t^{(1)}_i = t_{i+1}-t_i ,
\qquad
\Delta t^{(2)}_i = t_{i+2}-t_i .
\]
For the cluster size--waiting-time analysis, we further considered five directional waiting-time constructions derived from the same representative epochs. Their explicit definitions and physical interpretation are given in the main-text Results subsection ``Cluster size and waiting times''.

\subsection{Gaussian consistency tests, model comparisons and correlation analysis}
To assess Gaussian consistency of the clustered waiting-time sequences, we used the standard Anderson--Darling (AD) empirical-distribution-function statistic for the normal family. Unlike the Cram\'er--von Mises statistic, which applies uniform weighting across the cumulative distribution, the AD statistic weights deviations by \([F(x)(1-F(x))]^{-1}\) and is therefore more sensitive to discrepancies in the tails.

Because the Gaussian parameters were estimated from the data, the nominal critical values were supplemented by Monte Carlo calibration. Synthetic samples were drawn from the fitted normal model, the mean and standard deviation were re-estimated for each realization, and the AD statistic was recomputed to obtain the empirical null distribution and Monte Carlo \(p\)-value. Unless stated otherwise, \(10^4\) Monte Carlo realizations were used. We further evaluated robustness by excluding the maximum, the minimum, or both extremes, and by leave-one-out recalculation of the AD statistic.

To compare whether the clustered waiting-time sequences require a preferred timescale, we also fitted Gaussian, bounded uniform and exponential models. For each model, parameters were estimated by maximum likelihood, and the fits were compared using the log-likelihood, AICc and BIC. Goodness of fit was further evaluated by parametric-bootstrap versions of the Kolmogorov--Smirnov (KS), Anderson--Darling (AD) and Cram\'er--von Mises (CvM) statistics, in which synthetic samples were generated from the fitted model, the same model was re-fitted, and the statistic was recalculated to obtain the corresponding null distribution.

For the cluster size--waiting-time relations, we report Pearson, Spearman and Kendall correlation coefficients. Because the sample is small and some relations are sensitive to high-leverage points, significance was assessed by permutation calibration rather than by asymptotic formulae. Bootstrap resampling was used to estimate 95\% confidence intervals on the correlation coefficients.

\subsection{Treatment of the unresolved latest interval}
The two small glitches reported in July and August 2025 satisfy the temporal criterion for a cluster, but the cumulative amplitude of the latest cluster remains unresolved because no accompanying major glitch has yet been confirmed. Accordingly, in correlation analyses involving cluster size, relations that would otherwise depend on the unknown final cluster amplitude were evaluated after conservatively excluding the unresolved endpoint.

For timing constructions that require closure of the current open interval following the 2017 large glitch, we additionally considered a provisional zero-amplitude placeholder cluster located at the midpoint between the 2017 large glitch and the July--August 2025 activity. This working construction was used only to examine the consequences of a potentially unresolved skipped interval in the timing analysis; it should not be interpreted as the detection of an additional physical glitch.

\section*{Data availability}
The Crab glitches data used in this study were obtained from the Jodrell Bank Observatory pulsar glitch catalog\footnote{\url{http://www.jb.man.ac.uk/pulsar/glitches.html}}\cite{basu2022jodrell}. 
\section*{Code availability}
All codes written for the analysis are available from the
authors upon request.

\bibliographystyle{scibull}
\bibliography{references}

@article{ho2012rotational,
  title={Rotational evolution of young pulsars due to superfluid decoupling},
  author={Ho, Wynn CG and Andersson, Nils},
  journal={Nature Physics},
  volume={8},
  number={11},
  pages={787--789},
  year={2012},
  publisher={Nature Publishing Group UK London}
}

@article{page2011rapid,
  title={Rapid Cooling of the Neutron Star in Cassiopeia A Triggered by Neutron Superfluidity in Dense Matter},
  author={Page, Dany and Prakash, Madappa and Lattimer, James M and Steiner, Andrew W},
  journal={Physical Review Letters},
  volume={106},
  number={8},
  pages={081101},
  year={2011},
  publisher={APS}
}

@article{vivekanand2017non,
  title={A non-glitch speed-up event in the Crab Pulsar},
  author={Vivekanand, M},
  journal={Astronomy \& Astrophysics},
  volume={597},
  pages={L9},
  year={2017},
  publisher={EDP Sciences}
}

@article{zhu2024glitches,
  title={Glitches and glitching clusters in rotation-powered pulsars},
  author={Zhu, Pei-Xin and Zheng, Xiao-Ping},
  journal={The Astrophysical Journal},
  volume={978},
  number={1},
  pages={49},
  year={2024},
  publisher={IOP Publishing}
}

@article{eya2017angular,
  title={Angular momentum transfer and fractional moment of Inertia in pulsar glitches},
  author={Eya, IO and Urama, JO and Chukwude, AE},
  journal={The Astrophysical Journal},
  volume={840},
  number={1},
  pages={56},
  year={2017},
  publisher={IOP Publishing}
}

@article{fulgenzi2017radio,
  title={Radio pulsar glitches as a state-dependent Poisson process},
  author={Fulgenzi, W and Melatos, A and Hughes, BD},
  journal={Monthly Notices of the Royal Astronomical Society},
  volume={470},
  number={4},
  pages={4307--4329},
  year={2017},
  publisher={Oxford University Press}
}

@article{bak1987self,
  title={Self-organized criticality: An explanation of the 1/f noise},
  author={Bak, Per and Tang, Chao and Wiesenfeld, Kurt},
  journal={Physical review letters},
  volume={59},
  number={4},
  pages={381},
  year={1987},
  publisher={APS}
}

@incollection{antonelli2023insights,
  title={Insights into the physics of neutron star interiors from pulsar glitches},
  author={Antonelli, Marco and Montoli, Alessandro and Pizzochero, Pierre M},
  booktitle={Astrophysics in the XXI Century with Compact Stars},
  pages={219--281},
  year={2023},
  publisher={World Scientific}
}

@article{antonopoulou2022pulsar,
  title={Pulsar glitches: observations and physical interpretation},
  author={Antonopoulou, Danai and Haskell, Brynmor and Espinoza, Crist{\'o}bal M},
  journal={Reports on Progress in Physics},
  volume={85},
  number={12},
  pages={126901},
  year={2022},
  publisher={IOP Publishing}
}

@article{carlin2019temporal,
  title={Temporal clustering of rotational glitches in the Crab pulsar},
  author={Carlin, Julian Brian and Melatos, Andrew and Vukcevic, Damjan},
  journal={Monthly Notices of the Royal Astronomical Society},
  volume={482},
  number={3},
  pages={3736--3743},
  year={2019},
  publisher={Oxford University Press}
}

@article{ge2020discovery,
  title={Discovery of delayed spin-up behavior following two large glitches in the Crab pulsar, and the statistics of such processes},
  author={Ge, MY and Zhang, SN and Lu, FJ and Li, TP and Yuan, JP and Zheng, XP and Huang, Y and Zheng, SJ and Chen, YP and Chang, Z and others},
  journal={The Astrophysical Journal},
  volume={896},
  number={1},
  pages={55},
  year={2020},
  publisher={IOP Publishing}
}

@article{lyne1992spin,
  title={Spin-up and recovery in the 1989 glitch of the Crab pulsar},
  author={Lyne, AG and Smith, F Graham and Pritchard, RS},
  journal={Nature},
  volume={359},
  number={6397},
  pages={706--707},
  year={1992},
  publisher={Nature Publishing Group UK London}
}

@article{howitt2018nonparametric,
  title={Nonparametric estimation of the size and waiting time distributions of pulsar glitches},
  author={Howitt, G and Melatos, A and Delaigle, A},
  journal={The Astrophysical Journal},
  volume={867},
  number={1},
  pages={60},
  year={2018},
  publisher={IOP Publishing}
}

@article{middleditch2006predicting,
  title={Predicting the starquakes in PSR J0537--6910},
  author={Middleditch, John and Marshall, Francis E and Wang, Q Daniel and Gotthelf, Eric V and Zhang, William},
  journal={The Astrophysical Journal},
  volume={652},
  number={2},
  pages={1531},
  year={2006},
  publisher={IOP Publishing}
}

@article{basu2022jodrell,
  title={The Jodrell bank glitch catalogue: 106 new rotational glitches in 70 pulsars},
  author={Basu, Avishek and Shaw, Benjamin and Antonopoulou, Danai and Keith, Michael J and Lyne, Andrew G and Mickaliger, Mitchell B and Stappers, Benjamin W and Weltevrede, Patrick and Jordan, Christine A},
  journal={Monthly Notices of the Royal Astronomical Society},
  volume={510},
  number={3},
  pages={4049--4062},
  year={2022},
  publisher={Oxford University Press}
}

@article{lyne201545,
  title={45 years of rotation of the Crab pulsar},
  author={Lyne, AG and Jordan, CA and Graham-Smith, Francis and Espinoza, CM and Stappers, BW and Weltevrede, P},
  journal={Monthly Notices of the Royal Astronomical Society},
  volume={446},
  number={1},
  pages={857--864},
  year={2015},
  publisher={Oxford University Press}
}

@article{haskell2015models,
  title={Models of pulsar glitches},
  author={Haskell, Brynmor and Melatos, Andrew},
  journal={International Journal of Modern Physics D},
  volume={24},
  number={03},
  pages={1530008},
  year={2015},
  publisher={World Scientific}
}

@article{melatos2008avalanche,
  title={Avalanche dynamics of radio pulsar glitches},
  author={Melatos, A and Peralta, Carlos and Wyithe, JSB},
  journal={The Astrophysical Journal},
  volume={672},
  number={2},
  pages={1103},
  year={2008},
  publisher={IOP Publishing}
}

@article{antonopoulou2018pulsar,
  title={Pulsar spin-down: the glitch-dominated rotation of PSR J0537- 6910},
  author={Antonopoulou, Danai and Espinoza, Cristobal M and Kuiper, Lucien and Andersson, Nils},
  journal={Monthly Notices of the Royal Astronomical Society},
  volume={473},
  number={2},
  pages={1644--1655},
  year={2018},
  publisher={Oxford University Press}
}

@article{link1999pulsar,
  title={Pulsar constraints on neutron star structure and equation of state},
  author={Link, Bennett and Epstein, Richard I and Lattimer, James M},
  journal={Physical Review Letters},
  volume={83},
  number={17},
  pages={3362},
  year={1999},
  publisher={APS}
}

@article{zheng2025persistent,
  title={The persistent shift in spin-down rate following the largest Crab pulsar glitch rules out external torque variations due to starquakes},
  author={Zheng, Xiao-Ping and Wang, Wei-Hua and Huang, Chun and Yuan, Jian-Ping and Yuan, Sheng-Jie},
  journal={The Astrophysical Journal},
  volume={982},
  number={2},
  pages={181},
  year={2025},
  publisher={IOP Publishing}
}

\section*{Author contributions}
Pei-Xin Zhu and Xiao-Ping Zheng conceived the underlying idea and model for this study and wrote the manuscript. Quan Cheng, Chenghui Niu and Erbil G\"{u}gercino\u{g}lu edited the manuscript and contributed to writing.

\section*{Competing interests}
The authors declare no competing interests.

\section*{Acknowledgments}
This work is supported by the
National Nature Science Foundation of China under grant Nos.12033001 and 12473039 and the
National SKA Program of China under grant No.2020SKA0120300.

\newpage
\clearpage        
\onecolumn          
\appendix          
\section{Supplementary}
\subsection{Data and Cluster Construction}
\label{s1}

To facilitate full reproducibility, Table~\ref{t1} provides the complete mapping from the original glitch sequence to the derived glitch clusters used throughout this work. For each cluster, we list the constituent glitch IDs, the full set of epochs and amplitudes, the size-weighted representative epoch, and the total cluster size. All quantities in Table~\ref{t1} were constructed following the procedure defined in the main-text Methods, and the same definitions were used consistently in all subsequent analyses.

\begin{table}[!htbp]
\centering
\caption{Summary of glitch clusters, including constituent glitches, epochs, and derived quantities. All values are given in years, and $\Delta\nu/\nu$ is in units of $10^{-9}$.}
\label{t1}
\resizebox{\textwidth}{!}{
\begin{tabular}{c c l l c c}
\toprule
Group & Glitch IDs & All epochs (yr) & All $\Delta\nu/\nu$ & Epoch$_{\rm weighted\,mean}$ & $\sum \Delta\nu/\nu$ \\
			\midrule
			1 & 1 & \{0.00\} & \{6.00\} & 0.00 & 6.00 \\
			2 & 2 & \{3.02\} & \{81.00\} & 3.02 & 81.00 \\
			3 & 3 & \{6.25\} & \{4.20\} & 6.25 & 4.20 \\
			4 & 4--8 & \{9.19, 9.85, 10.39, 10.48, 11.36\} & \{2.10, 31.90, 6.10, 0.80, 6.20\} & 10.10 & 47.10 \\
			5 & 9--15 & \{13.11, 13.90, 14.08, 14.84, 15.01, 15.97, 16.22\} & \{6.80, 25.10, 3.50, 22.60, 8.90, 3.40, 1.70\} & 14.42 & 72.00 \\
			6 & 16--18 & \{17.53, 18.04, 18.25\} & \{214.00, 4.90, 2.80\} & 17.55 & 221.70 \\
			7 & 19 & \{20.00\} & \{21.80\} & 20.00 & 21.80 \\
			8 & 20 & \{21.67\} & \{4.70\} & 21.67 & 4.70 \\
			9 & 21 & \{25.22\} & \{39.00\} & 25.22 & 39.00 \\
			10 & 22--26 & \{30.60, 31.21, 31.69, 32.33, 32.92\} & \{2.20, 516.37, 4.08, 2.30, 31.70\} & 31.32 & 556.65 \\
			11 & 27--28 & \{38.90, 38.96\} & \{2.33, 4.53\} & 38.94 & 6.86 \\
			\bottomrule
\end{tabular}}
\end{table}

\clearpage
\newpage
\subsection{Robustness of Gaussian consistency in the waiting-time distributions}
\label{s2}
To assess the impact of small-number statistics and tail sensitivity, we performed a series of robustness tests on two complementary waiting-time sequences: the primary sequence, $\Delta t^{(1)}$, defined from adjacent glitch-cluster intervals, and the alternative sequence, $\Delta t^{(2)}$, constructed from non-consecutive (every-other) waiting times.

To assess Gaussian consistency of the clustered waiting-time sequences, we used the standard Anderson--Darling (AD) empirical-distribution-function statistic for the normal family. Unlike the Cram\'er--von Mises statistic, which applies uniform weighting across the cumulative distribution, the AD statistic weights deviations by $[F(x)(1-F(x))]^{-1}$ and is therefore more sensitive to discrepancies in the tails. In practice, we computed the normal AD statistic using the standard implementation for the Gaussian family. Because the Gaussian parameters were estimated from the sample itself, we further calibrated the test by Monte Carlo: synthetic samples were drawn from the fitted normal model, the mean and standard deviation were re-estimated for each realization, and the AD statistic was recomputed to obtain the empirical null distribution and corresponding Monte Carlo $p$-value.

For $\Delta t^{(1)}$, the observed Anderson--Darling (AD) statistic is 0.35, which lies well within the Monte Carlo null distribution generated under the fitted Gaussian model (mean $\approx 0.36$; 95\% interval $\approx [0.15,\,0.79]$), yielding an empirical p-value of $\approx 0.42$. This indicates that the observed deviation from normality is consistent with statistical fluctuations expected under the null hypothesis. To examine the role of extreme observations, we repeated the analysis after removing the maximum, the minimum, and both extremes. In all cases, the AD statistic remained below the corresponding 5\% critical threshold, although the strongest effect is seen when the minimum is excluded, for which the AD statistic increases to $\approx 0.52$ and the empirical Monte Carlo p-value decreases to $p \approx 0.14$. Thus, the qualitative inference is unchanged under plausible modifications of the tail structure. We further quantified sensitivity to individual observations using a leave-one-out analysis. For the original $\Delta t^{(1)}$ sequence, the AD statistic varies over a limited range ($\approx 0.24$--$0.52$) and remains below the critical value in all realizations. The Monte Carlo calibrated KS p-values remain above 0.05 throughout, although smaller values occur for the most tail-sensitive configurations. These results indicate that the non-rejection of normality for $\Delta t^{(1)}$ is not driven by any single data point, while showing modest sensitivity to the lower-tail structure.
	
For $\Delta t^{(2)}$, consistency with the Gaussian model is stronger. The observed AD statistic is 0.16, again well within the corresponding Monte Carlo null distribution, with an empirical p-value of $\approx 0.95$. Tail-modified analyses yield similarly small AD statistics, all remaining below the corresponding 5\% critical threshold, together with consistently high empirical p-values. The largest change is obtained after excluding the minimum, for which the AD statistic increases only to $\approx 0.21$ and the empirical Monte Carlo p-value decreases to $p \approx 0.82$, without altering the qualitative inference. A leave-one-out analysis of the original $\Delta t^{(2)}$ sequence further supports this result: the AD statistic varies only modestly ($\approx 0.14$--$0.21$), and no realization approaches the corresponding critical threshold. The Monte Carlo calibrated KS p-values also remain uniformly high across all realizations. These results indicate that the Gaussian-compatible inference for $\Delta t^{(2)}$ is robust against both extreme-value treatment and single-point perturbation.
	
As an additional consistency check, we also evaluated a Kolmogorov--Smirnov-type statistic with Monte Carlo calibration. This test is broadly consistent with the AD-based inference for both datasets. In particular, the KS results remain uniformly high for $\Delta t^{(2)}$, while for $\Delta t^{(1)}$ somewhat smaller values appear in the most tail-sensitive cases, as expected for a very small sample with slight tail imbalance. This does not alter the overall statistical interpretation. Taken together, these analyses show that, although $\Delta t^{(1)}$ exhibits modest sensitivity to tail structure, the inference of non-rejection of normality remains robust. The $\Delta t^{(2)}$ sequence provides an independent and more stable check, remaining statistically consistent with a Gaussian model under both tail-modified and leave-one-out analyses.

\begin{figure*}
\centering
\includegraphics[width=1.0\linewidth]{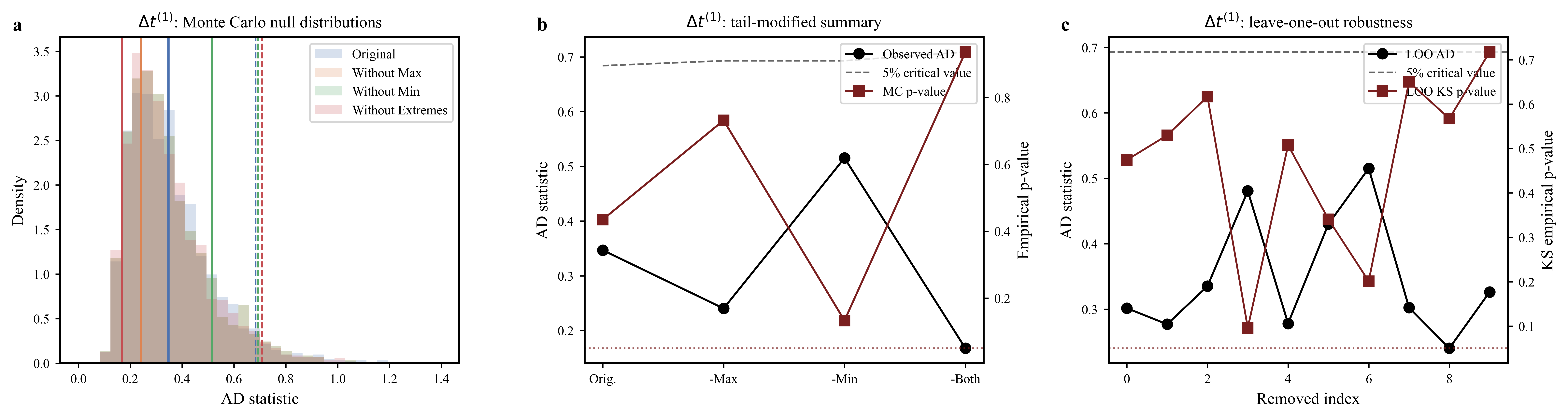}
\caption{\textbf{| Robustness tests for Gaussian consistency of the primary waiting-time sequence $\Delta t^{(1)}$.}
\textbf{a}, Monte Carlo null distributions of the Anderson--Darling (AD) statistic under the fitted Gaussian model for the original dataset and for datasets obtained after excluding the maximum, the minimum, and both extremes. Solid vertical lines mark the observed AD statistics, and dashed vertical lines indicate the corresponding 5\% critical values. For the original sequence, the observed AD statistic (0.35) lies well within the null distribution and below the 5\% critical threshold, consistent with the empirical Monte Carlo p-value ($p \approx 0.42$). The same qualitative conclusion is retained after all tail-modified treatments.
\textbf{b}, Summary of the observed AD statistic and empirical Monte Carlo p-value for the four tail-modified datasets. Excluding the minimum produces the largest AD statistic ($\sim 0.52$) and the lowest empirical p-value ($p \approx 0.14$), indicating that the lower tail contributes most strongly to the residual shape sensitivity. However, all AD statistics remain below the corresponding 5\% critical values, and all empirical p-values remain non-significant.
\textbf{c}, Leave-one-out robustness analysis for the original $\Delta t^{(1)}$ sequence. The AD statistic varies over a limited range ($\sim 0.24$--$0.52$) and remains below the corresponding 5\% critical value for all realizations. The Monte Carlo calibrated KS p-value remains above 0.05 in all cases, although a lower value appears in the most tail-sensitive configuration. Together, these results show that the Gaussian-compatible inference for $\Delta t^{(1)}$ is not driven by any single observation and is only modestly sensitive to tail structure.}
\end{figure*}
	
\begin{figure*}
\centering
\includegraphics[width=1.0\linewidth]{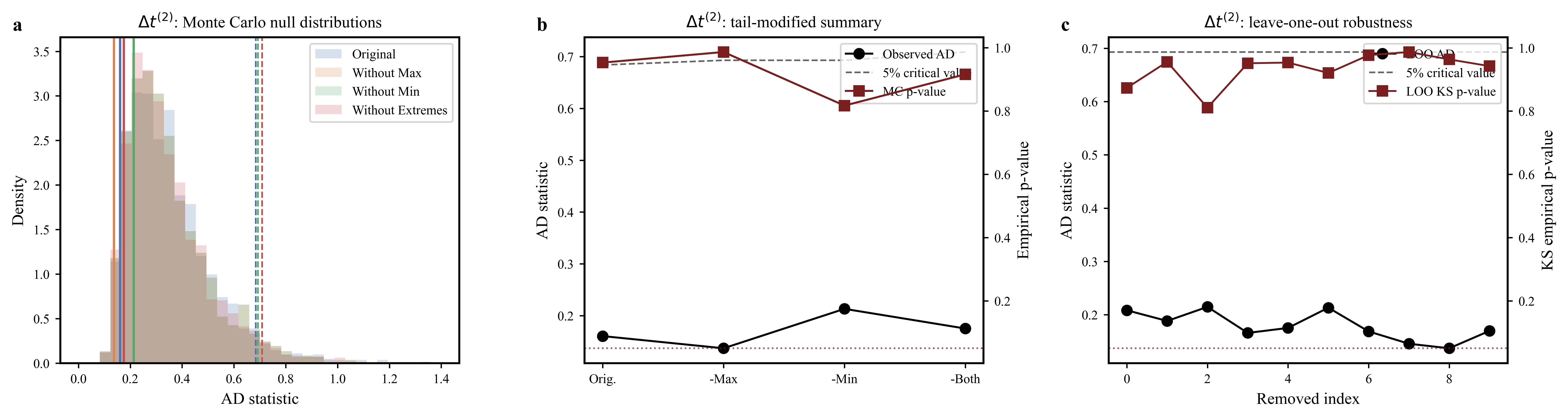}
\caption{\textbf{| Robustness tests for Gaussian consistency of the alternative waiting-time sequence $\Delta t^{(2)}$.}
\textbf{a}, Monte Carlo null distributions of the Anderson--Darling (AD) statistic under the fitted Gaussian model for the original dataset and for datasets obtained after excluding the maximum, the minimum, and both extremes. Solid vertical lines mark the observed AD statistics, and dashed vertical lines indicate the corresponding 5\% critical values. For the original sequence, the observed AD statistic (0.16) lies well within the null distribution and far below the 5\% critical threshold, consistent with the empirical Monte Carlo p-value ($p \approx 0.95$). The same qualitative conclusion is retained after all tail-modified treatments.
\textbf{b}, Summary of the observed AD statistic and empirical Monte Carlo p-value for the four tail-modified datasets. All AD statistics remain small ($\lesssim 0.21$) and all empirical p-values remain high ($p \gtrsim 0.82$), indicating weak sensitivity to extreme-value treatment.
\textbf{c}, Leave-one-out robustness analysis for the original $\Delta t^{(2)}$ sequence. The AD statistic varies only over a narrow range ($\sim 0.14$--$0.21$) and remains below the corresponding 5\% critical value for all realizations. The Monte Carlo calibrated KS p-value also remains uniformly high across all realizations, showing that the Gaussian-compatible inference is not driven by any individual observation.}
\end{figure*}
\clearpage
\newpage
\clearpage

\subsection{Model comparison for the clustered waiting-time sequences}
\label{s3}
To examine whether the clustered waiting-time sequences require a Gaussian marginal distribution, we compared three simple families for both the consecutive waiting times, $\Delta t^{(1)}$, and the every-other waiting times, $\Delta t^{(2)}$: a Gaussian model, a bounded uniform model, and an exponential model. The Gaussian model represents a distribution with a characteristic central scale, whereas the bounded uniform model serves as a non-peaked benchmark over a finite interval. The exponential model represents the expectation under an idealized memoryless waiting-time process.

For each model, we estimated the parameters by maximum likelihood and summarized the fits using the log-likelihood, AICc, BIC, and parametric-bootstrap goodness-of-fit tests based on the Kolmogorov--Smirnov (KS), Anderson--Darling (AD), and Cram\'er--von Mises (CvM) statistics (Supplementary Table~\ref{t2}). In the bootstrap procedure, synthetic samples were generated from the fitted model, the same model was re-fitted to each realization, and the corresponding test statistic was recalculated to construct the null distribution. The empirical cumulative distribution functions together with the Gaussian and bounded uniform fits are shown in Supplementary Fig.~\ref{f7}.

For both $\Delta t^{(1)}$ and $\Delta t^{(2)}$, the exponential model is strongly disfavoured by both the information criteria and the bootstrap goodness-of-fit tests. By contrast, neither the Gaussian nor the bounded uniform model is rejected at the present sample size. For $\Delta t^{(1)}$, the two models are descriptively close, indicating that the marginal distribution alone does not clearly distinguish between a peaked and a non-peaked finite-support description. For $\Delta t^{(2)}$, the bounded uniform model yields smaller AICc and BIC values than the Gaussian model, although this difference should be interpreted cautiously because the support parameters of the bounded uniform model are estimated from the sample extrema.

Overall, these comparisons show that Gaussian consistency alone does not uniquely identify a preferred timescale in the clustered waiting-time data. Our inference of quasi-periodic temporal structure therefore does not rest on marginal distribution fitting alone, but on the combined evidence from the clustered $\sim 3.5$-yr spacing, the corresponding $\sim 7$-yr every-other spacing, and the dedicated midpoint Monte Carlo analysis presented in Supplementary Note~\ref{s6}.

\clearpage
\newpage
\clearpage

\begin{figure*}
	\centering
	\includegraphics[width=0.95\linewidth, height=0.4\textheight, keepaspectratio]{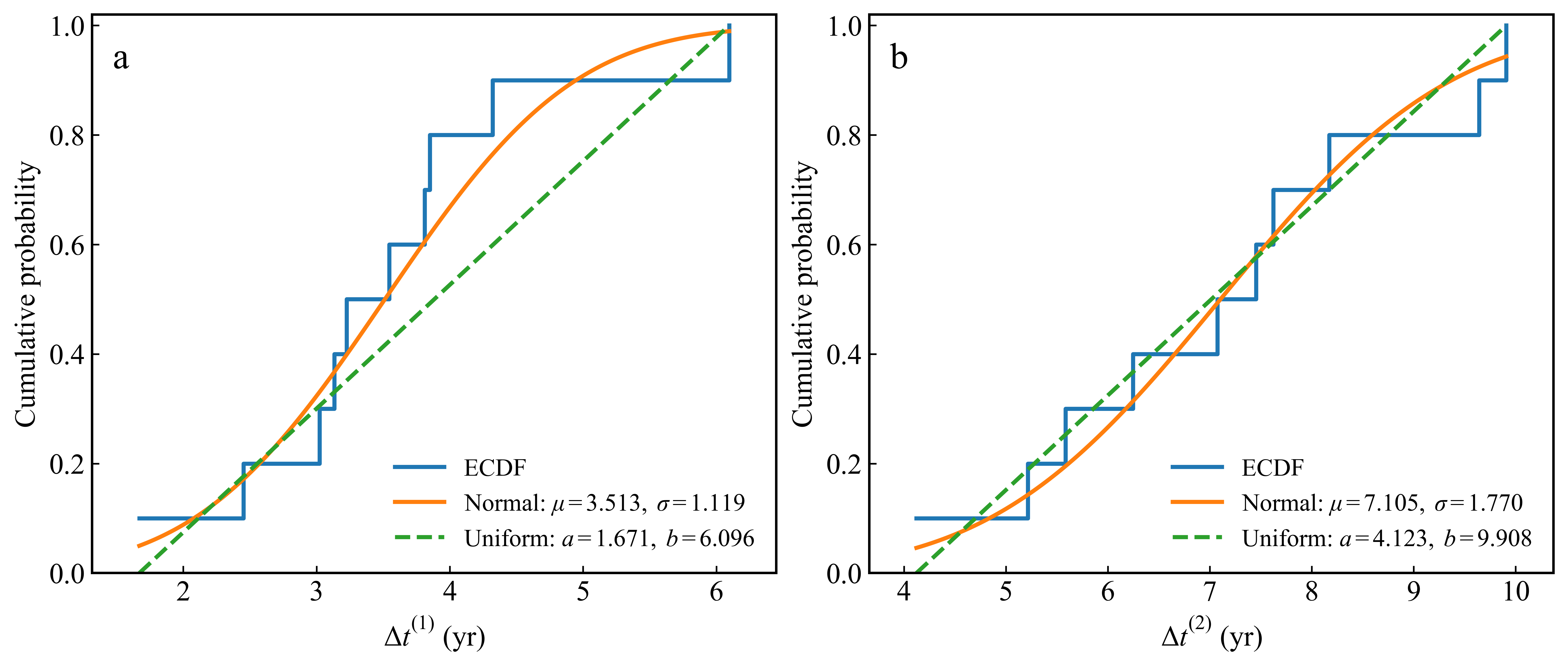}
	\caption{\textbf{| Comparison of empirical waiting-time distributions with Gaussian and bounded uniform models.}
	\textbf{a}, Empirical cumulative distribution function (ECDF) of the consecutive cluster waiting times, $\Delta t^{(1)}$, together with the maximum-likelihood Gaussian and bounded uniform fits.
	\textbf{b}, Same as in \textbf{a}, but for the every-other waiting times, $\Delta t^{(2)}$.
	In both cases, the bounded uniform model cannot be excluded at the level of marginal distribution fitting, indicating that Gaussian consistency alone does not uniquely establish a preferred timescale. The exponential fit is omitted for clarity because it is strongly disfavoured by both the information criteria and the bootstrap goodness-of-fit tests.}
\label{f7}
\end{figure*}

\begin{table*}
	\centering
    \begin{threeparttable}
	\caption{\textbf{|Model comparison and parametric-bootstrap goodness-of-fit tests for the clustered waiting-time sequences.}}

	\setlength{\tabcolsep}{5pt}
	\renewcommand{\arraystretch}{1.15}
	\begin{tabular}{lcccccccc}
		\toprule
		Model & MLE & logL & AICc & $\Delta$AICc & BIC & KS $p_{\rm boot}$ & AD $p_{\rm boot}$ & CvM $p_{\rm boot}$ \\
		\midrule
		\multicolumn{9}{l}{$\Delta t^{(1)}$} \\
		Uniform     & $a=1.671,\ b=6.096$        & $-14.874$ & $35.46$ & $0.00$  & $34.35$ & $0.244$  & $0.529$  & $0.277$  \\
		Normal      & $\mu=3.513,\ \sigma=1.119$ & $-15.312$ & $36.34$ & $0.88$  & $35.23$ & $0.489$  & $0.512$  & $0.568$  \\
		Exponential & $\lambda=0.285$            & $-22.564$ & $47.63$ & $12.17$ & $47.43$ & $0.0051$ & $0.0049$ & $0.0014$ \\
		\midrule
		\multicolumn{9}{l}{$\Delta t^{(2)}$} \\
		Uniform     & $a=4.123,\ b=9.908$        & $-17.553$ & $40.82$ & $0.00$  & $39.71$ & $0.957$  & $0.930$  & $0.979$  \\
		Normal      & $\mu=7.105,\ \sigma=1.770$ & $-19.899$ & $45.51$ & $4.69$  & $44.40$ & $0.950$  & $0.944$  & $0.966$  \\
		Exponential & $\lambda=0.141$            & $-29.608$ & $61.72$ & $20.89$ & $61.52$ & $0.0020$ & $0.0030$ & $0.0007$ \\
		\bottomrule
	\end{tabular}%
    \begin{tablenotes}[flushleft]
	\footnotesize
	\item Maximum-likelihood estimates (MLEs), information criteria, and bootstrap goodness-of-fit $p$-values are shown for the consecutive waiting times, $\Delta t^{(1)}$, and the every-other waiting times, $\Delta t^{(2)}$. For each dataset, $\Delta$AICc is defined relative to the model with the smallest AICc. Information-criterion differences involving the bounded uniform model should be interpreted cautiously, because its support parameters are estimated from the sample extrema.
\end{tablenotes}
\end{threeparttable}
\label{t2}
\end{table*}

\clearpage
\newpage
\clearpage
\subsection{Supplementary analysis of the remaining waiting-time constructions}

To test whether the cluster-size correlations emphasized in the main text are selective rather than generic, we examined the three remaining waiting-time constructions not shown there: the post-cluster, mid-two-step and post-two-step cases. These supplementary comparisons show that the cluster-based correlation pattern is highly non-uniform across constructions. The post-cluster relation shows an apparent linear trend, but this is not supported by the rank-based statistics and is therefore sensitive to leverage from the largest cluster. The post-two-step relation likewise fails to show a comparably robust positive association. The mid-two-step case also exhibits a visually suggestive trend, but again the rank-based coefficients remain substantially weaker than the Pearson statistic. In contrast to the pre-two-step result emphasized in the main text, these three cases do not provide equally strong or internally consistent evidence for longer-timescale directional dependence.

This asymmetry is physically suggestive. If the observed correlations were a generic by-product of regrouping, one might expect similarly strong behaviour across several waiting-time constructions. Instead, the signal remains concentrated in the preceding intervals, especially in the longer pre-two-step case. The supplementary comparisons therefore support the interpretation that the cluster-size relation is directional and history-dependent, with the clearest imprint carried by the $\sim 7$-yr pre-two-step timescale. The full numerical results, including permutation-calibrated $p$-values and bootstrap 95\% confidence intervals, are given in Supplementary Table~\ref{t3}.

\begin{table*}
	\centering
	\small
	\setlength{\tabcolsep}{4pt}
	\renewcommand{\arraystretch}{1.15}
	\begin{tabular}{l c c c p{2.3cm} c c p{2.3cm} c c p{2.3cm}}
		\toprule
		& & \multicolumn{3}{c}{Pearson} & \multicolumn{3}{c}{Spearman} & \multicolumn{3}{c}{Kendall} \\
		\cmidrule(lr){3-5}\cmidrule(lr){6-8}\cmidrule(lr){9-11}
		Case & $N$ & $r$ & $p_{\rm perm}$ & 95\% CI & $\rho$ & $p_{\rm perm}$ & 95\% CI & $\tau$ & $p_{\rm perm}$ & 95\% CI \\
		\midrule
		Post          & 10 &  0.626 & 0.0892 & [-0.595, 0.944] &  0.103 & 0.7924 & [-0.711, 0.733] &  0.022 & 1.0000 & [-0.588, 0.590] \\
		Pre           &  9 &  0.785 & 0.0298 & [-0.180, 0.959] &  0.483 & 0.1994 & [-0.232, 1.000] &  0.444 & 0.1240 & [-0.125, 1.000] \\
		Post-two-step &  9 & -0.451 & 0.1930 & [-0.877, 0.237] & -0.517 & 0.1638 & [-0.945, 0.214] & -0.389 & 0.1862 & [-0.862, 0.161] \\
		Mid-two-step  &  9 &  0.734 & 0.0649 & [-0.539, 0.954] &  0.383 & 0.3127 & [-0.426, 0.842] &  0.167 & 0.6118 & [-0.500, 0.677] \\
		Pre-two-step  &  8 &  0.787 & 0.0086 & [ 0.453, 0.971] &  0.810 & 0.0195 & [ 0.200, 1.000] &  0.643 & 0.0317 & [ 0.040, 1.000] \\
		\bottomrule
	\end{tabular}
\caption{\textbf{| Correlation statistics for cluster size versus five waiting-time constructions.}
Pearson, Spearman and Kendall correlation coefficients are listed for the post, pre, post-two-step, mid-two-step and pre-two-step waiting times, together with permutation-calibrated $p$-values and bootstrap 95\% confidence intervals. Sample sizes differ because of the indexing of the five constructions and the conservative exclusion of relations involving the unresolved latest cluster where appropriate. The table shows that the strongest and most internally consistent positive signal is obtained for the pre-two-step relation, whereas the pre relation is weaker but remains directionally consistent. By contrast, the post-cluster and post-two-step relations do not show comparably robust support, and the mid-two-step trend is substantially more sensitive to the choice of statistic. These results support the main conclusion that the cluster-size correlations are directional and history-dependent rather than generic across arbitrary waiting-time definitions.}
\label{t3}
\end{table*}

\begin{figure*}
\centering
\includegraphics[width=0.95\linewidth]{s4}
\caption{\textbf{| Scatter plots for the three non-primary waiting-time constructions.}
\textbf{a}, Cluster size versus post-cluster waiting time. An apparent linear trend is present, but the corresponding rank-based statistics are weak, indicating that the relation is sensitive to leverage and does not provide robust support for a post-event correlation.
\textbf{b}, Cluster size versus mid-two-step waiting time. A visually suggestive trend is visible, but it is not matched by equally strong rank-based coefficients, indicating limited robustness and enhanced sensitivity to the placement of high-leverage points.
\textbf{c}, Cluster size versus post-two-step waiting time. No comparably robust positive association is found on this longer post-event timescale; several correlation measures are instead weak or negative. Together, these three cases contrast with the pre and pre-two-step relations highlighted in the main text, showing that the cluster-size correlations are neither temporally symmetric nor generic across all waiting-time constructions.}
\label{f8}
\end{figure*}

\clearpage
\newpage
\clearpage

\subsection{Illustrative extrapolations for possible future larger activity}
\label{s5}
Following the prospective consistency discussed in the main text, we used the updated adjacent and pre-two-step size--waiting-time relations to obtain illustrative extrapolations for a possible larger future event. These extrapolations are intended only to show how the present waiting-time history maps onto the range of cluster sizes represented in the current sample.

Such estimates must be interpreted cautiously. The cumulative amplitude of the latest cluster remains unresolved, and the available sample is small. We therefore do not treat the fitted relations as deterministic forecasting laws, but only as heuristic mappings between the observed waiting-time history and the cluster sizes already present in the clustered Crab sequence.

Using the adjacent pre-cluster relation,
$$ \Delta \nu_{-9}/\nu = 112.02\, t_{\mathrm{pre,yr}}^{(1)} - 273.33 $$
where $\Delta \nu_{-9}$ is in units of $10^{-9}$ and $t^{(1)}_{\mathrm{pre,yr}}$ is in years, the elapsed time of 8.05~yr since the latest major glitch gives an extrapolated cluster size of
$$ \Delta \nu/\nu = 628.4 \times 10^{-9}, $$
with a 95\% confidence interval of [255.5, 1001.4]

Using the longer-baseline pre-two-step relation,
$$\Delta \nu_{-9}/\nu = 84.44\, t_{\mathrm{pre,yr}}^{(2)} - 443.98,$$
the elapsed time of 14.04~yr since the penultimate major glitch gives a corresponding extrapolated size of
$$\Delta \nu/\nu = 741.6 \times 10^{-9},
$$
with a 95\% confidence interval of [242.6, 1240.6].

The broad consistency between the adjacent and pre-two-step extrapolations indicates that, within the updated cluster-based framework, the recent July--August 2025 activity is compatible with an unresolved active episode whose eventual cumulative amplitude could exceed that of the two observed minor glitches alone. We emphasize, however, that these values are heuristic rather than predictive in a strict statistical sense, and are included here only to illustrate the implications of the fitted size--waiting-time relations under the precursor interpretation discussed in the main text.

\clearpage
\newpage
\clearpage

\subsection{Monte Carlo test of the midpoint-coincidence interpretation of the apparent \texorpdfstring{$\sim 3.5$}{~3.5}-yr pattern}
\label{s6}
A specific alternative explanation for the apparent $\sim 3.5$\,yr spacing is that it does not reflect an intrinsic characteristic timescale of the clustered glitch sequence, but instead arises because smaller glitch clusters happen to lie near the midpoints between successive large glitch clusters. To test this conditional null hypothesis, we fixed the observed epochs of the large clusters and redistributed the smaller clusters uniformly within the relevant inter-large-cluster intervals, while preserving the observed multiplicity in each interval.

We classified the 2nd, 4th, 6th, 9th, 10th, and 11th clusters in the sequence as the large glitch clusters, with epochs
\[
[3.02,\ 10.10,\ 17.55,\ 25.22,\ 31.32,\ 38.94]\ {\rm yr},
\]
and treated the remaining clusters as smaller clusters. These large clusters define five successive inter-large-cluster intervals. Only the first three intervals contain one or more smaller clusters, so the present test is restricted to the subset of intervals in which midpoint concentration can be meaningfully defined. This limitation should be kept in mind when interpreting the result.

For each interval containing at least one smaller cluster, we defined the interval midpoint $m_i$ and the normalized nearest-midpoint distance
$$d_i=\frac{\min_j |t_{ij}-m_i|}{L_i/2},$$
where $t_{ij}$ is the epoch of the $j$th smaller cluster in interval $i$, and $L_i$ is the corresponding interval length. By construction, $d_i=0$ corresponds to exact coincidence with the midpoint and $d_i=1$ corresponds to the interval boundary. This normalization places all relevant intervals on the same scale and allows direct comparison despite their differing lengths.

For the three relevant intervals, the observed values are
$$d_i=[0.088,\ 0.159,\ 0.075].$$
We further summarize the overall degree of midpoint concentration using
$$D_{\rm mean}=\frac{1}{N}\sum_i d_i,
\qquad D_{\max}=\max_i d_i, $$
which for the observed data give
$$D_{\rm mean}=0.108, \qquad D_{\max}=0.159.$$
We then constructed a null model in which the large-cluster epochs were held fixed, while the smaller clusters were redistributed uniformly at random within each of the three relevant inter-large-cluster intervals, preserving the observed number of smaller clusters in each interval. For each Monte Carlo realization, we recalculated $d_i$, $D_{\rm mean}$, and $D_{\max}$. We generated $2\times10^5$ such realizations.

Under this null model, the probability of obtaining a value of $D_{\rm mean}$ at least as small as observed is
$ p_{\rm mean}=0.010 $,
and the probability of obtaining a value of $D_{\max}$ at least as small as observed is
$ p_{\max}=0.0075. $
We also evaluated a stricter joint criterion, namely the probability that all three relevant intervals are at least as concentrated toward their respective midpoints as observed, and obtained
$ p_{\rm all}=0.0020. $

Taken together, these results indicate that, within the currently relevant inter-large-cluster intervals, the observed midpoint concentration is less readily explained by uniform random placement than by a more structured arrangement. At the same time, the inference remains limited by the small number of relevant intervals and by the absence of comparable midpoint-associated activity in all inter-large-cluster intervals. This test therefore does not establish an independent $\sim 3.5$\,yr periodicity. Rather, it shows that the apparent midpoint alignment is more structured than expected under a simple random-placement null hypothesis, and thus weakens the most direct midpoint-coincidence explanation of the clustered pattern.
\begin{figure}
	\centering
	\includegraphics[width=0.95\linewidth]{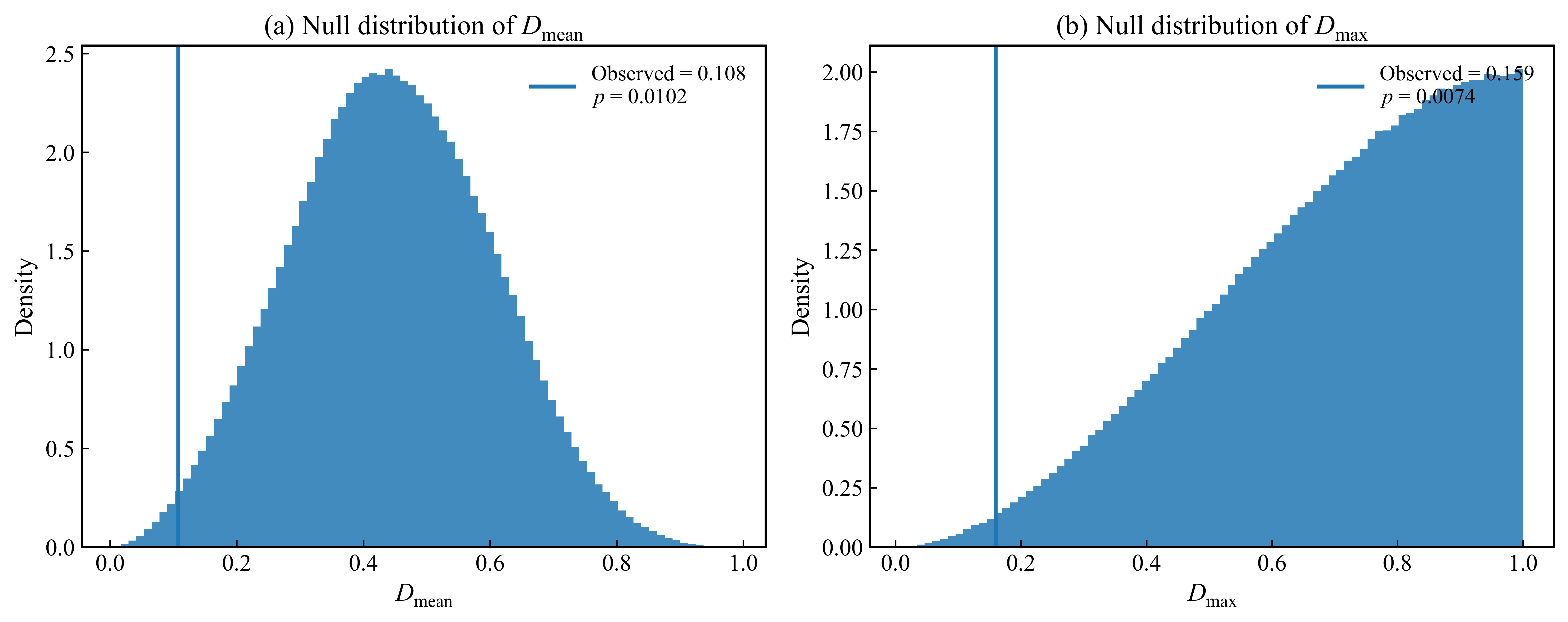}
	\caption{\textbf{| Monte Carlo null distributions of the midpoint-concentration statistics for the smaller glitch clusters.}
		\label{f9}
		The epochs of the large glitch clusters were held fixed, and the smaller clusters were redistributed uniformly within the three inter-large-cluster intervals that contain one or more smaller clusters, while preserving the observed number of smaller clusters in each interval. For each Monte Carlo realization, we computed the normalized nearest-midpoint distances $d_i$ and the summary statistics $D_{\rm mean}=\frac{1}{N}\sum_i d_i$ and $D_{\max}=\max_i d_i$.
		\textbf{a}, Null distribution of $D_{\rm mean}$. The observed value, $D_{\rm mean}=0.108$, is marked by the vertical line and lies in the lower tail of the null distribution, with Monte Carlo probability $p_{\rm mean}=0.010$.
		\textbf{b}, Null distribution of $D_{\max}$. The observed value, $D_{\max}=0.159$, is likewise marked by the vertical line and yields $p_{\max}=0.0075$. For the three relevant intervals, the observed normalized midpoint distances are $d_i=[0.088,\,0.159,\,0.075]$, and the joint probability that all intervals are at least as concentrated toward their midpoints as observed is $p_{\rm all}=0.0020$. These results indicate that, within the currently relevant intervals, the observed midpoint concentration is less readily explained by uniform random placement than by a more structured arrangement.}
\end{figure}
\end{document}